\documentclass{sig-alternate-05-2015}

\usepackage{balance}  % to better equalize the last page
\usepackage{graphics} % for EPS, load graphicx instead
\usepackage{times}    % comment if you want LaTeX's default font
\usepackage{url}      % llt: nicely formatted URLs
\usepackage{longtable}

\usepackage{caption}
\usepackage{subcaption}

\makeatletter
\def\url@leostyle{%
  \@ifundefined{selectfont}{\def\UrlFont{\sf}}{\def\UrlFont{\small\bf\ttfamily}}}
\makeatother
\def\pprw{8.5in}
\def\pprh{11in}

\usepackage[pdftex]{hyperref}
\hypersetup{
pdftitle={SIGCHI Conference Proceedings Format},
pdfauthor={LaTeX},
pdfkeywords={SIGCHI, proceedings, archival format},
bookmarksnumbered,
pdfstartview={FitH},
colorlinks,
citecolor=black,
filecolor=black,
linkcolor=black,
urlcolor=black,
breaklinks=true,
}

\begin{document}

\title{Gaze-Assisted User Authentication to Counter Shoulder-surfing Attacks}

\numberofauthors{1}
\author{
  \alignauthor Vijay Rajanna, Dr. Tracy Hammond\\
    \affaddr{Sketch Recognition Lab. Dept. of Computer Science and Engineering}\\
    \affaddr{Texas A\&M University}\\
    \email{vijay.drajanna@gmail.com, thammond@gmail.com}\\
  }

\maketitle

%%%%%%%%%%%%%%%%%%%%%%%%%%%%%%%%%%%%%%%%%%%%%%%%%%%%%%%%%%%%%%%%%%%%%%%%%%%%%%%%%%%%%%%%%%%%%%%%%%%
% GOALS
%%%%%%%%%%%%%%%%%%%%%%%%%%%%%%%%%%%%%%%%%%%%%%%%%%%%%%%%%%%%%%%%%%%%%%%%%%%%%%%%%%%%%%%%%%%%%%%%%%%
% The technical problem to be solved with a justification of its importance
% An account of related and prior works that explains why these works have not solved the problem
% The specific research problem or question that your thesis work will address
% A sketch of the proposed approach or solution
% The expected contributions of your dissertation research
% Progress in solving the stated problem
% The methods you are using or will use to carry out your research
% A plan for evaluating your work and presenting credible evidence of your results to the research community

%%%%%%%%%%%%%%%%%%%%%%%%%%%%%%%%%%%%%%%%%%%%%%%%%%%%%%%%%%%%%%%%%%%%%%%%%%%%%%%%%%%%%%%%
\begin{abstract}
A highly secure, foolproof, user authentication system is still a primary focus of research in the field of User Privacy and Security.
Shoulder-surfing is an act of spying when an authorized user is logging into a system, and is promoted by a malicious intent of gaining unauthorized access.
We present a gaze-assisted user authentication system as a potential solution to counter shoulder-surfing attacks.
The system comprises of an eye tracker and an authentication interface with 12 pre-defined shapes (e.g., triangle, circle, etc.) that move simultaneously on the screen.
A user chooses a set of three shapes as a password.
To authenticate, the user follows the paths of three shapes as they move, one on each frame, over three consecutive frames.

The system uses either the template matching or decision tree algorithms to match the scan-path of the user's gaze with the path traversed by the shape.
The system was evaluated with seven users to test the accuracy of both the algorithms.
We found that with the template matching algorithm the system achieves an accuracy of 95\%, and with the decision tree algorithm an accuracy of 90.2\%.
We also present the advantages and disadvantages of using both the algorithms.
Our study suggests that gaze-based authentication is a highly secure method against shoulder-surfing attacks as the unique pattern of eye movements for each individual makes the system hard to break into.
\end{abstract}



%%%%%%%%%%%%%%%%%%%%%%%%%%%%%%%%%%%%%%%%%%%%%%%%%%%%%%%%%%%%%%%%%%%%%%%%%%%%%%%%%%%%%%%%
\section{Introduction}
The need for secure user authentication is gaining a significant interest as the number of unauthorized accesses are growing.
Unauthorized access allows attackers to gain access to the private and confidential information of users.
In general, attackers gain unauthorized access using two means:  1) brute force attacks, 2) shoulder-surfing attacks.
Brute force is an unintelligent attack where the person gaining unauthorized access tries a trial and error method to guess the right password.
This is a time-consuming process and demands patience. 
However, brute force attacks have a higher rate of success when the attacker has no constraints on the lifetime of the password.
Since brute force attack tries all the possible permutations and combinations of the potential password, the search becomes comprehensive leading to a higher time demand.

On the contrary, shoulder-surfing is a direct observation technique in which the attacker looks over a victim's shoulder to get the confidential information\footnote{http://searchsecurity.techtarget.com/definition/shoulder-surfing}.
Shoulder-surfing attacks are prevalent in crowded places as the victim is hardly able to recognize a potential attacker.
In most of the cases an attacker tries to capture personal identification information of the victim to gain unauthorized access.
In addition, the attackers can take help of advanced technologies and vision-enhancing devices like a long range binocular, thermal camera, and so on to support their attacks.
Hence a reliable system to effectively prevent shoulder-surfing attacks is much needed.

There has been a considerable number of solutions proposed to counter shoulder-surfing attacks, specifically in public places like ATMs.
Most of these solutions use gaze as the primary input modality to enter the secured information, e.g., a password~\cite{Kumar:2007:RSU:1280680.1280683, DeLuca:2009:LME:1572532.1572542}.
These systems, however, suffer from low accuracy issues and induce high cognitive load.
In addition, these solutions expect the gaze input to be highly precise enough to operate (click) keys on a virtual keyboard, or draw specific gestures with gaze by connecting the dots on a grid.
When using these solutions if the gaze input is not precise, the system functionality becomes unreliable.

In this work we take a new approach to gaze-based user authentication.
Here, instead of asking the user to draw a pattern or input keys using gaze, we ask the user to follow an object on the screen.
To authenticate, a user has to follow 3 objects, one at a time, over three consecutive frames.
The objects have a predefined shape, an onscreen position, color, and most importantly a definite path.
The user chooses 3 shapes as a password, and authenticates herself by following the path of the three shapes.
For a successful authentication, the scan-path of the user's gaze should match with the path traversed by the object on all the three frames.

For matching the user's scan-path to the path traversed by the object we use two algorithms: 1) template matching, and 2) decision tree.
Template matching algorithm matches the user's scan-path with a predefined template path that is at the nearest distance to the path traversed with the gaze.
On the other hand, decision tree based system extracts the key features of template paths, and generates a model to match against the user's scan-path.
To match the user's scan-path with a template path, first, the key features of the scan-path are extracted and the feature values are processed through a decision tree.
The outcome of the decision tree assigns a final class (shape) to the user's scan-path.
During the initial development phase, the system was evaluated for it's accuracy in recognizing scan-paths generated from a total of 7 participants.
We found that though the template matching algorithm is more accurate than the decision tree algorithm, it consumes more time in recognizing the scan-paths.

%%%%%%%%%%%%%%%%%%%%%%%%%%%%%%%%%%%%%%%%%%%%%%%%%%%%%%%%%%%%%%%%%%%%%%%%%%%%%%%%%%%%%%%%%%
\section{Prior Work}
While gaze input has been successfully used for point-and-click interactions~\cite{Rajanna:gawschi, Rajanna:IUI}, text entry~\cite{Rajanna:foottyping}, and activity recognition~\cite{Kaul:sketch}, we discuss some of the gaze-based solutions to address shoulder-surfing attacks.
Bulling et al.~\cite{Bulling:2012:ISG:2207676.2208712}, presented a novel gaze-based authentication system that makes use of cued-recall graphical passwords on a single image.
The algorithm leverages a computational model of visual attention to mask those areas of the image that are obvious spots of visual attention.
Based on a 12-participants user study, the authors showed that their algorithm is significantly more secure than a standard image-based authentication and gaze-based 4-digit pin entry.
Specifically, the authors used a bottom-up computational model of visual saliency that aims at estimating the parts of a visual scene that are most likely to attract visual attention.
To set a password, the authors present to the user an image with all parts of the image masked that are more likely to draw the user's attention.

De Luca et al.~\cite{DeLuca:2007:EEI:1324892.1324932}, evaluated three different eye gaze interaction methods for PIN entry specifically designed to be resistant against shoulder-surfing attacks.
The authors also investigated a new approach of gaze gestures and compared it to the well known classical gaze-interactions.
The three modes of gaze interactions presented were 1) dwell time-based input method, 2) look and shoot method, and 3) gaze gestures.
Another work by De Luca et al. \cite{DeLuca:2009:LME:1572532.1572542}, presented a solution for an authentication system that is used in public terminals.
The authors used an authentication method ``EyePassShapes," that uses eye gestures.
EyePassShapes combines and extends the two authentication methods: 1) PassShapes, and 2) EyePIN .
The PassShapes authentication uses a series of strokes as the PIN.
The series of strokes are encoded into a string, for example, ``URDL" represents-Up, Right, Down, and Left.
In the EyePIN method, the user performs eye gestures that represent the respective digits.
The system evaluation showed that the EyePassShapes can significantly increase security while being easy to use and fast at the same time.

Moncur et al.~\cite{Moncur:2007:PAE:1240624.1240758}, conducted multiple experiments to evaluate whether users find multiple graphical passwords more memorable than PIN-based passwords.
The results showed that multiple graphical passwords are substantially more effective than multiple PIN-based passwords.
The study presented two memory augmentation strategies.
Graphical passwords consisting of detailed, colorful, meaningful photographic images were assigned to the users.
Each participant was assigned 5 passwords to test the retention over a four week period.
The authors tested a total of 3 hypothesis: 1) memorability of multiple graphical passwords over PIN , 2) using mnemonics to aid recall , and 3) password and distractor images against a signature colored background.
The longitudinal study proved that graphical passwords can be recalled with higher success rate than the PIN passwords.

Kumar et al.~\cite{Kumar:2007:RSU:1280680.1280683}, presented ``EyePassword," a system that can mitigate shoulder-surfing by using novel input methods.
With EyePassword, a user enters a sensitive information by selecting from an on-screen keyboard by moving the gaze.
This is one of the earliest works in Gaze-based authentication systems.
The authors presented multiple design choices, their usability, and the level of security offered.
The system uses two trigger method: 1) Gaze+Dwell , and 2) Gaze+Trigger for key selection.
Dwell time based method is appropriate for triggering the key press since it does not reveal the timing information.
The results show that gaze-based password entry though takes relatively higher time, it performs as efficiently as keyboard-based input.
Also, QWERTY layout outperforms Alpha-numeric keyboard.

Roth et al.~\cite{Roth:2004:PMR:1030083.1030116}, presented an alternative PIN entry method using ``Cognitive Trapdoor Games."
This PIN entry method makes it extremely hard to break into the system.
In addition, the authors also introduced an idea of probabilistic cognitive trapdoor games, which fails attempt to recover the PIN by recording the PIN entry procedure using a camera.
The PIN entry process is designed like a game play, involving 3 participants: a machine interrogator, a human oracle, and a human observer.
Both the interrogator and the oracle receive a key from a dealer.
The human observer will observe the oracle as she tries to log-in.
The oracle authenticates herself to the system by answering questions presented by the system.
At each step the oracle is presented with two partitions colored black and white, and must input in which partition the current PIN digit is.

Though these solutions attempt to prevent shoulder-surfing, there are common limitations observed among them.
These solutions do not achieve the expected accuracy of an authentication system, induce cognitive load, and cannot be adopted in various scenarios.
Hence, a system that is accurate, induces no cognitive load, and can easily be adopted to multiple scenarios still remains unrealized.
In this work, we focus on building a gaze-assisted authentication system that prevents shoulder-surfing, while also addressing the limitations with the existing systems.

%%%%%%%%%%%%%%%%%%%%%%%%%%%%%%%%%%%%%%%%%%%%%%%%%%%%%%%%%%%%%%%%%%%%%%%%%%%%%%%%%%%%%%%%
\section{System Architecture}
\begin{figure}[!ht]
\centering
\includegraphics[width=0.8\columnwidth]{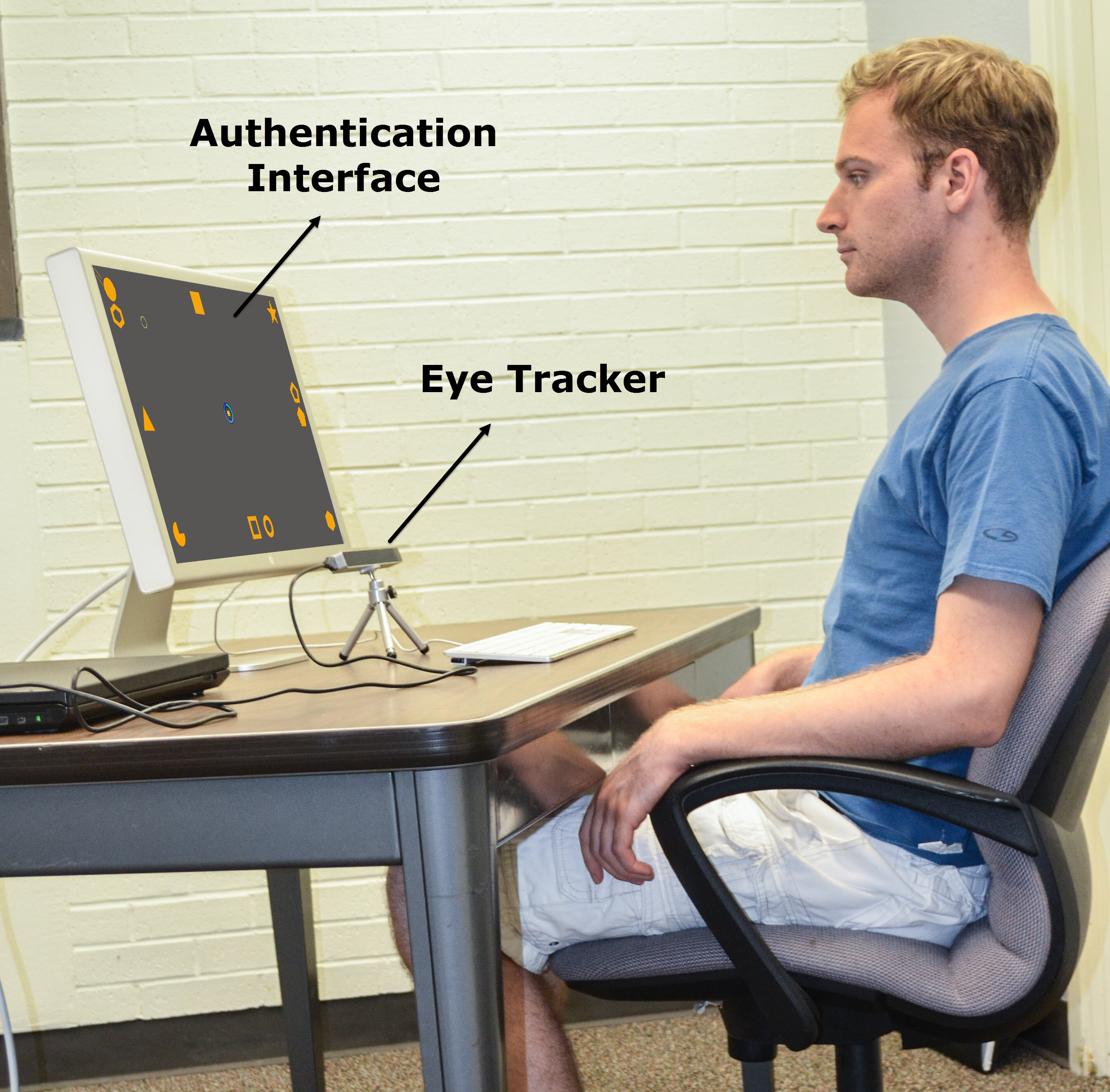}
\caption{A user authenticating himself by following the three shapes selected as the password}
\label{figure:user}
\end{figure}

The gaze-assisted authentication system consists of two main modules: 1) Gaze Tracking Module, and 2) Authentication Engine.
In addition, the the user interface has a minimal number of objects to achieve a sufficient complexity of the the passwords, but does not clutter the interface.
A working model of the system is depicted in Figure \ref{figure:user}.

%====================================================================
\subsection{Gaze Tracking Module}
The system uses the ``Eye Tribe" tracker\footnote{theeyetribe.com\label{eyetribe}}, which is a table mounted eye tracker that provides a pair of (X,Y) coordinates as the user's gaze moves on the screen.
For the eye tracker to work efficiently, the user's position is adjusted such that the face is centered in front of the monitor at a distance of 45 - 75 cm~\textsuperscript{\ref{eyetribe}}.
For each user the system needs to be calibrated to prevent gaze tracking errors while following the shapes with the gaze.
The authentication engine is the central module to which the eye tracking module starts streaming the data as the user looks at different points on the screen.

%====================================================================
\subsection{Authentication Engine}
Authentication engine implements both template matching and decision tree algorithms to recognize the user's scan-paths as as it receives gaze tracking data from the eye tracker.
The authentication engine has an interface with multiple shapes with each shape traversing along a specific path.

%%%%%%%%%%%%%%%%%%%%%%%%%%%%%%%%%%%%%%%%%%%%%%%%%%%%%%%%%%%%%%%%%%%%%%%%%%%%%%%%%%%
\section{Authentication Procedure}

%====================================================================
\subsection{Password Selection}
\begin{figure}[!ht]
\centering
\includegraphics[width=0.8\columnwidth]{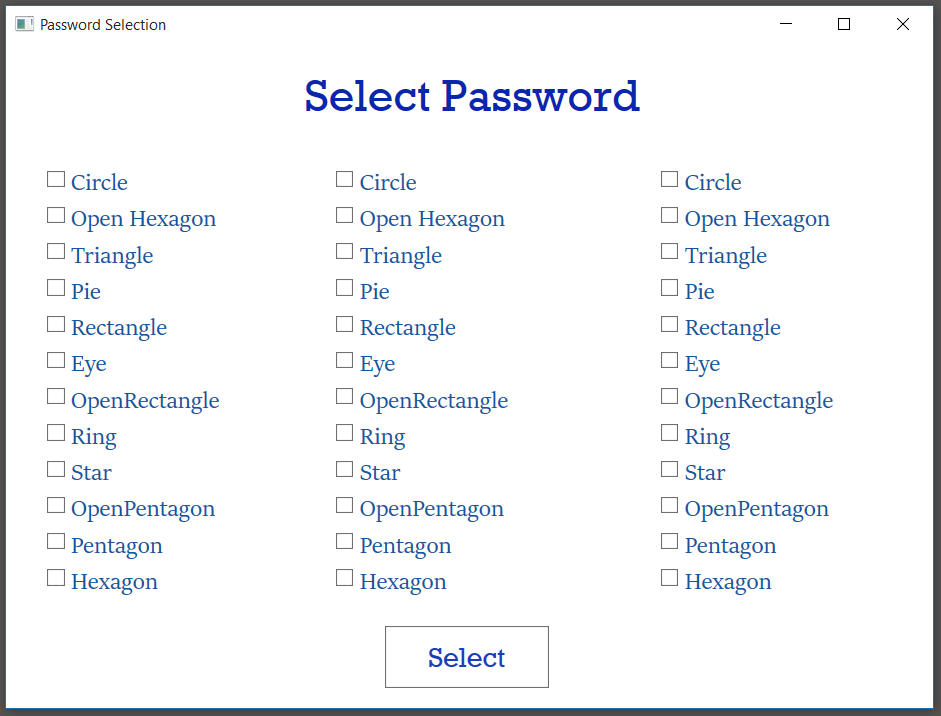}
\caption{Password selection dialog}
\label{figure:password}
\end{figure}

To authenticate, a user first selects 3 shapes from the password selection window shown in Figure \ref{figure:password}.
The three columns correspond to the three frames, and the user has to follow each shape selected on a column in its corresponding frame.
Password selection is a one time process, however if the user wants to change the password at any time, the same interface allows for changing the password.

%====================================================================
\subsection{Authentication Interface}
\begin{figure}[!ht]
\centering
\includegraphics[width=0.8\columnwidth]{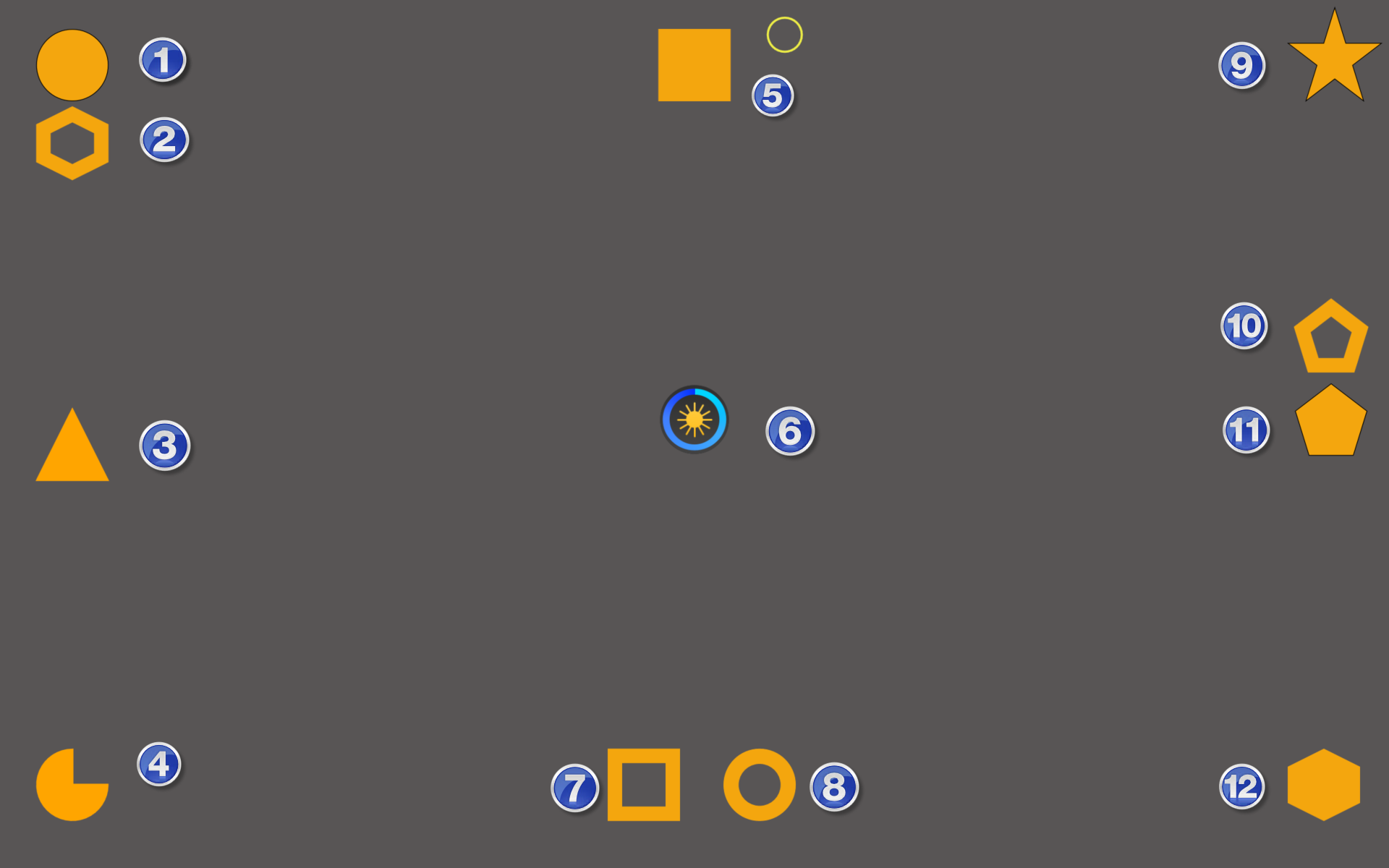}
\caption{Authentication interface}
\label{figure:interface}
\end{figure}

The authentication interface is shown in Figure \ref{figure:interface}.
The interface is a canvas with 12 shapes placed at different locations on the screen.
The beginning and ending positions of each shape are pre-defined.
Also, each object is assigned a pre-defined path along which the shape traverses when a frame is under execution.

%======================================================================================
\subsection{Password Entry}
\begin{figure}[!ht]
\centering
\includegraphics[width=0.8\columnwidth]{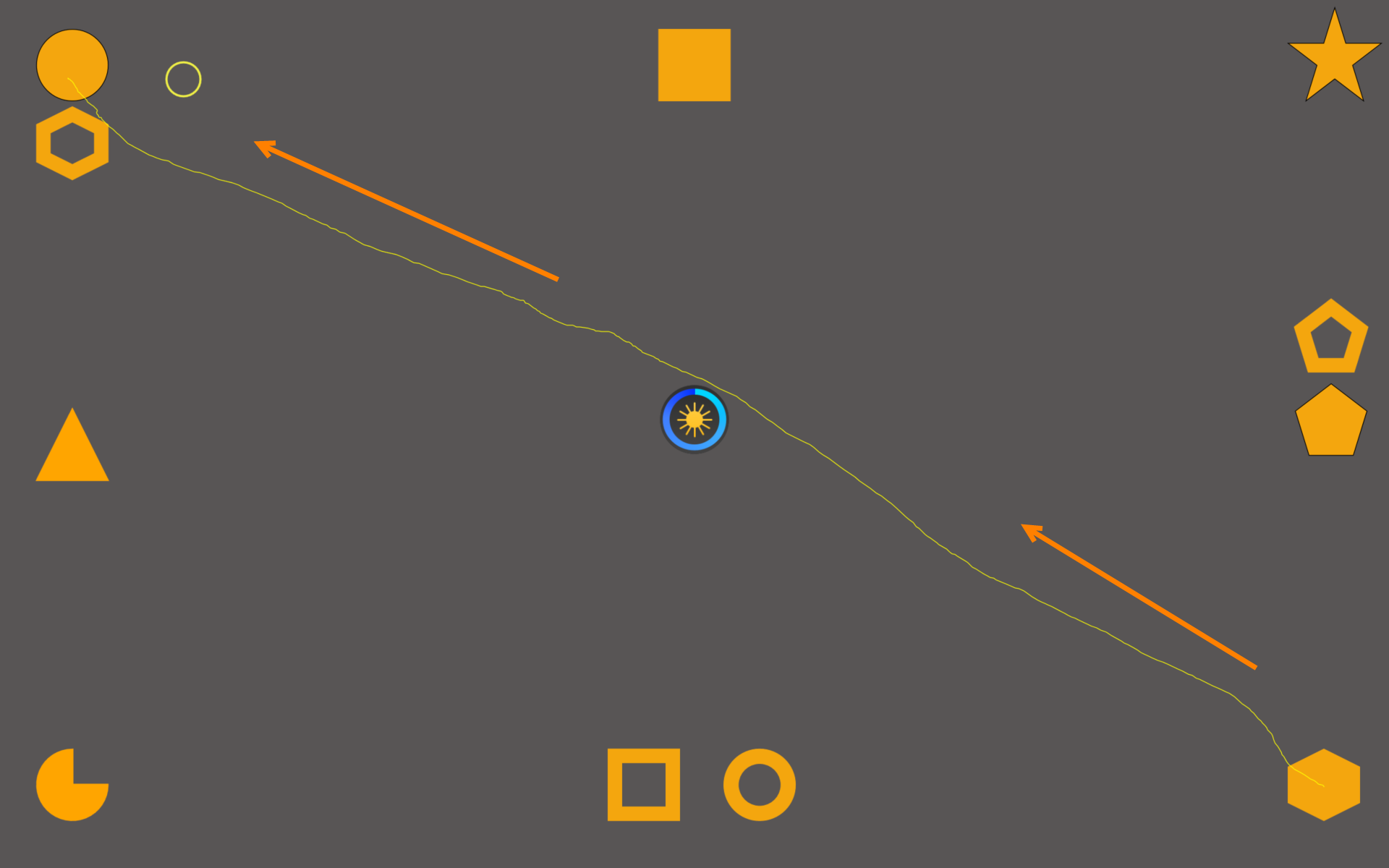}
\caption{Follow the hexagon on frame-1}
\label{figure:hexagon}
\end{figure}

\begin{figure}[!ht]
\centering
\includegraphics[width=0.8\columnwidth]{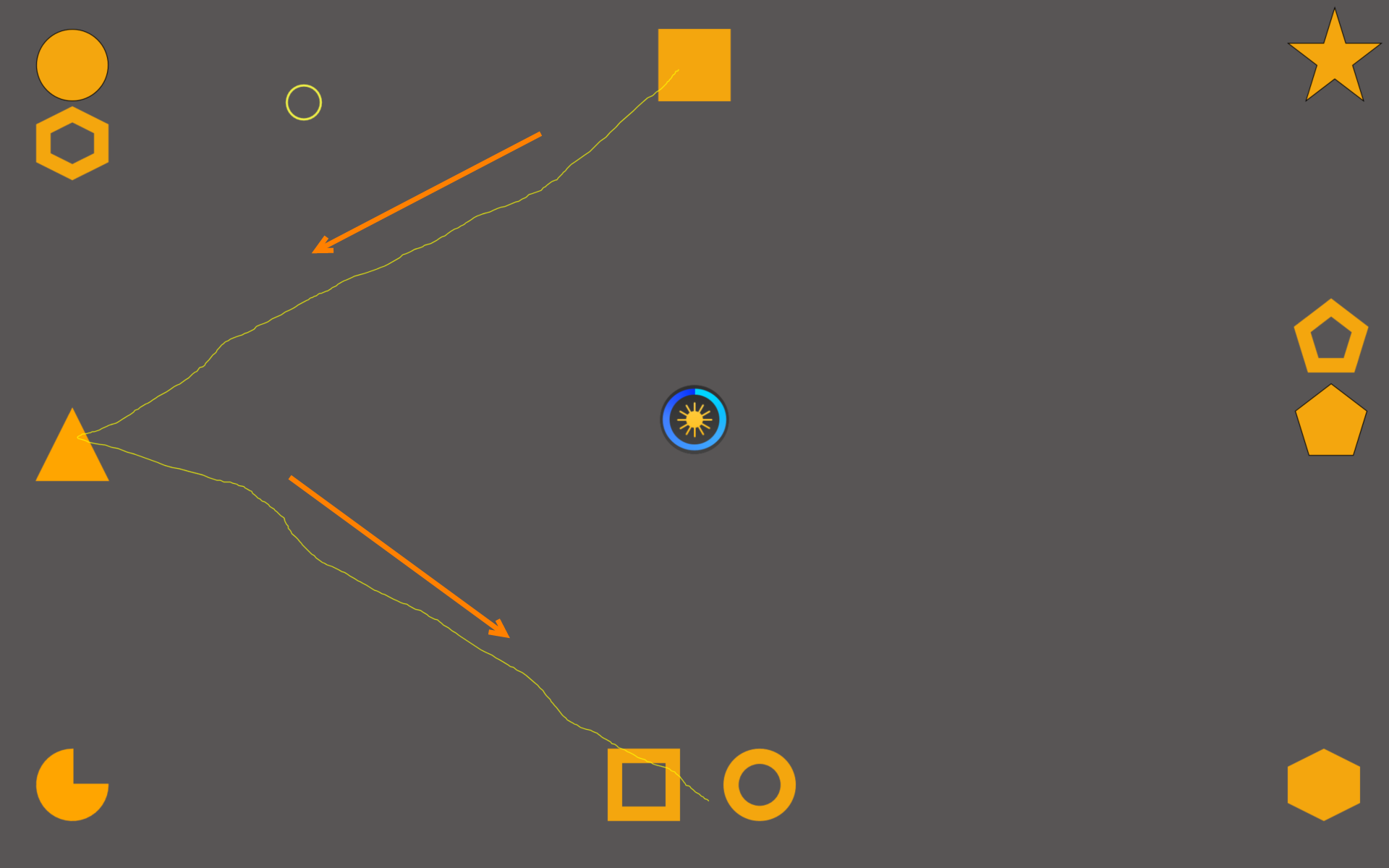}
\caption{Follow the rectangle on frame-1}
\label{figure:rectangle}
\end{figure}

\begin{figure}[!ht]
\centering
\includegraphics[width=0.8\columnwidth]{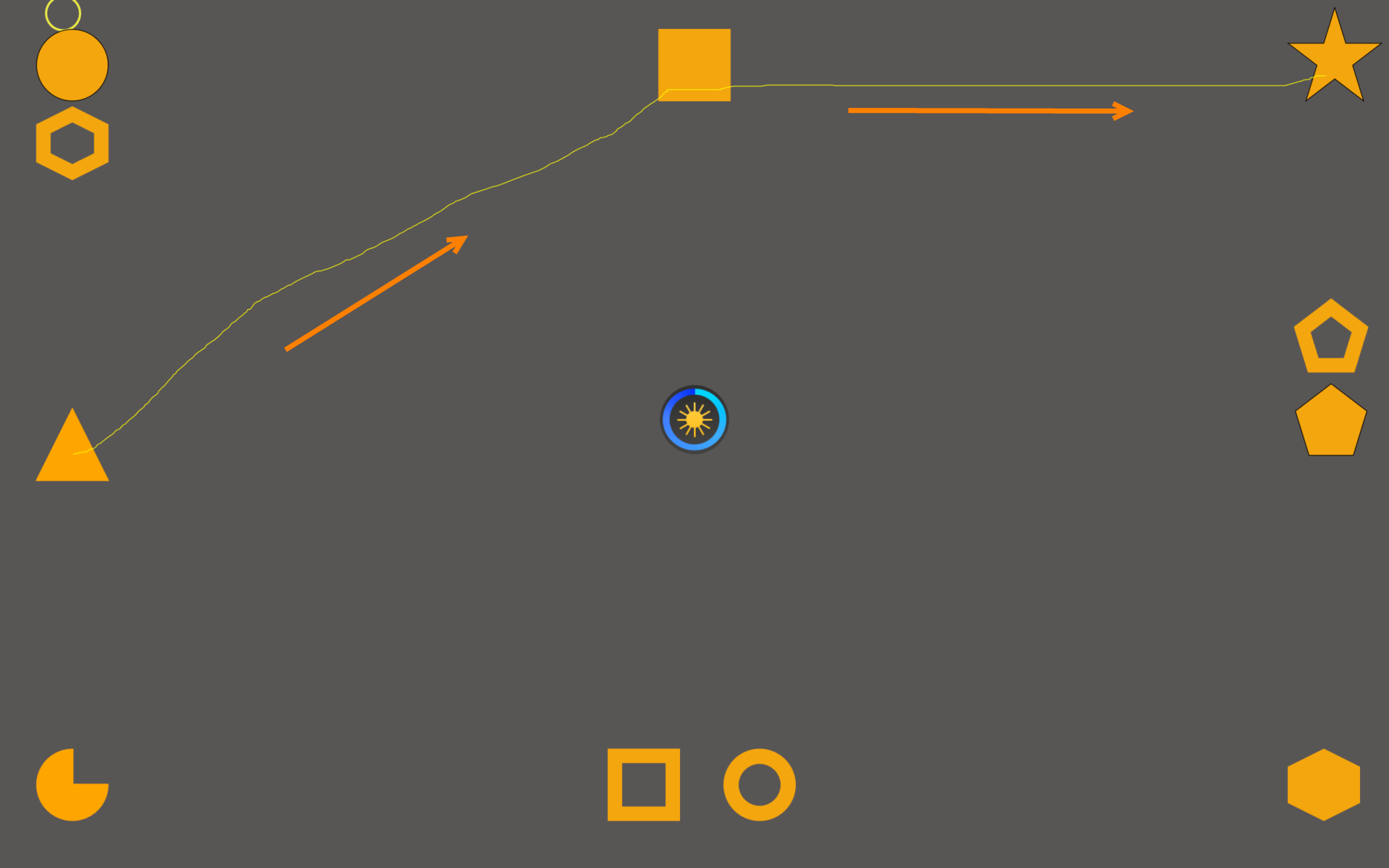}
\caption{Follow the triangle on frame-1}
\label{figure:triangle}
\end{figure}

A set of three shapes constitute a password, a user has to follow paths traversed by those three shapes to authenticate.
For example, if the user has selected Hexagon, Rectangle and Triangle as the password, when authenticating the user will follow those three shapes, in sequence, one on each frame.
On the first frame the user will follow the hexagon as shown in figure \ref{figure:hexagon}.
On the second frame the user will follow the rectangle as shown in figure \ref{figure:rectangle}.
Lastly, on the third frame the user will follow the triangle as shown in figure \ref{figure:triangle}.

%======================================================================================
\subsection{Training}
The system has a default set of template paths for each shape.
However, for a user specific model, the user can train the system by providing a new set of templates.
This can be achieved by following a shape on each frame, and repeating the same process for all the shapes on the screen.
By re-training, the system generates a new set of templates to be used by the template matching algorithm.
In addition, a model is also generated using the same templates for the decision tree algorithm.

% %======================================================================================
% \subsection{User Authentication}
% To authenticate, a user first initiates the eye tracker and makes sure that the system is calibrated.
% Further the user follows the selected 3 shapes, one on each frame, over three consecutive frames.
% The system validates shapes drawn by the user against the password selected. 
% If the path of the shapes match the user is authenticated.
% Otherwise the access is denied.

%%%%%%%%%%%%%%%%%%%%%%%%%%%%%%%%%%%%%%%%%%%%%%%%%%%%%%%%%%%%%%%%%%%%%%%%%%%%%%%%%%%%%%%%
\section{Recognition}
Matching a user's scan-path against a template is achieved through two algorithms: 1) template matching, and 2) decision tree.
We wanted the algorithms to be highly responsive, and either authenticate or deny the access with no delay.
Therefore, based on the performance of multiple algorithms we tested, we finally chose template matching and decision tree algorithms.

%======================================================================================
\subsection{Template Matching Algorithm}
The Template Matching algorithm compares two paths, and computes the distance of the candidate path (user's gaze-path) from the template path (shape's actual path).
When a candidate path matches closely with the path traversed by a shape, the shape is selected as the object followed by the user.
The template matching algorithm adopts most of the processing stages presented in \cite{Wobbrock:2007:GWL:1294211.1294238}, but each stage is modified to suit the needs of our application.
The path matching process involves Sampling, Scaling, Translation, and Recognition steps:

\subsubsection{Sampling}
In the sampling stage a path is sampled to 64 points that are distributed equally along the path.
To compute the offset between two points on a path, we first compute the total path length.
Next, starting with the first point we add the offset distance, and choose the closest point that is at a least distance from the other end point of the offset, and make minor adjustments.
After sampling, all the strokes will have an equal number of stroke points.

\subsubsection{Scaling}
In this stage, the path is scaled to a square of 300 x 300 by scaling along x and y axes.

\subsubsection{Translation}
In this stage, we find the centroid of the path and translate the whole path to the origin point (0,0).

\subsubsection{Recognition - Template Matching}
Once the candidate path is processed it is matched against all the templates using equation 1, where P is a (X,Y) point on the path, C - candidate path, T - template path, and DT - distance to template.
The template at a least Euclidean distance from the scan-path is assigned as the final class of the scan-path.
\begin{equation}
DT = \frac{\sum \limits^{N}_{p=1} { \sqrt{(C[p]_x - T[p]_x)^2 + (C[p]_y - T[p]_y)^2} }}{N}
\end{equation}

%======================================================================================
\subsection{Decision Tree Algorithm}
\begin{figure}[!ht]
\centering
\includegraphics[width=0.8\columnwidth]{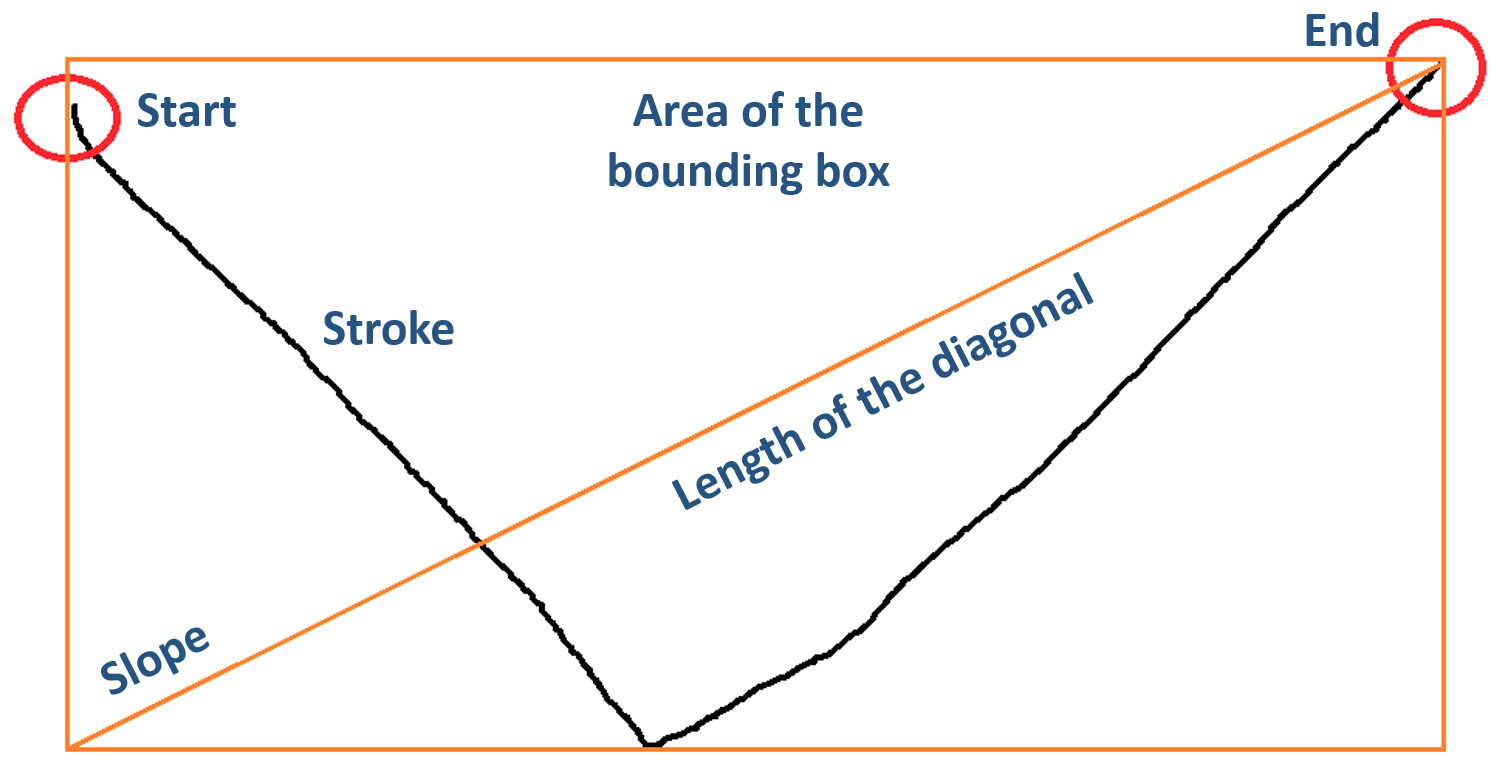}
\caption{Decision tree features}
\label{figure:decisiontree}
\end{figure}

Another approach toward recognizing the path traversed by the user is through the decision tree algorithm \cite{friedl1997decision}.
In this method, we first create the model for the classification algorithm using multiple template paths.
For creating the model, we extract the unique features of each path.
The features we considered are:
\begin{itemize}
\item Starting Point
\item Ending Point
\item Area of the bounding box
\item Length of the bounding box diagonal
\item Slope of the bounding box diagonal
\end{itemize}
A pictorial depiction of the features considered are shown in Figure \ref{figure:decisiontree}

%%%%%%%%%%%%%%%%%%%%%%%%%%%%%%%%%%%%%%%%%%%%%%%%%%%%%%%%%%%%%%%%%%%%%%%%%%%%%%%%%%%%%%%%
\section{Experiment Design and Results}
To test the accuracy of the recognition algorithms, we recruited 7 participants, and at the beginning of the experiment each participant was briefed about the experiment and calibrated with the eye tracking system.
Furthermore, each participant was asked to follow each shape twice resulting in the generation of two candidate scan-paths.
Data from one participant were discarded due to low tracking accuracy following the calibration.
To test the accuracy of the template matching algorithm, 10 scan-paths were selected randomly for each shape, making a total of 120 candidate paths (10 scan-paths x 12 shapes).
The template matching algorithm achieved an accuracy of 95\%.
Table \ref{table:confusion:template} shows the confusion matrix from the template matching algorithm.

a - Circle, b - Open Hexagon, c - Triangle, d - Pie, e - Rectangle, f- eye, g - Open rectangle, h - Ring, i - Start, j - Open pentagon, k - Pentagon, l - Hexagon
\setlength\tabcolsep{5.0pt}
\begin{table}[!ht]
\scriptsize
\caption{Confusion Matrix - Template Matching }
\label{table:confusion:template}
\begin{tabular}{|c|c|c|c|c|c|c|c|c|c|c|c|c|} \hline

&a&b&c&d&e&f&g&h&i&j&k&l\\ \hline

a&1.0 & & & & & & & & & & &  \\ \hline 
  
  b&& 0.8 & & & & & 0.2 & & & & &  \\ \hline
  
  c&&  & 0.7 & 0.3 & & & & & & & &  \\ \hline
 
  d&&  & & 1.0 & & & & & & & &  \\ \hline
 
  e&& 0.1 & & & 0.9 & & & & & & &  \\ \hline
  
  f&& & & & & 0.9 & 0.1 & & & & &  \\ \hline
  
  g&& & & & & & 0.9 & & &  0.1 & &  \\ \hline  
  
  h&& & & & & &  & 0.9 & & 0.1 & &  \\ \hline
  
  i&& & & & & &  &  & 1.0 & & &  \\ \hline
  
  j&& & & & & &  &  & & 0.9 & & 0.1  \\ \hline
  
  k&& & & & & &  &  & 0.2 & 0.1 & 0.7 &  \\ \hline
  
  l&& & & & & &  &  & & & & 1.0  \\ \hline   

\end{tabular}
\end{table}

To test the accuracy of the decision tree algorithm, we randomly selected one scan-path for each shape from each participant.
Hence, a total of 72 scan-paths were selected from six participants (6 participants * 12 shapes).
The decision tree algorithm achieved an accuracy of 90.2\%, and the confusion matrix generated from a 10 fold cross validation is shown in table \ref{table:confusion:decision}.
% For classification using decision tree we used Accord .NET machine learning framework\footnote{accord-framework.net}.
\begin{table}[!ht]
\scriptsize
\caption{Confusion Matrix - Decision Tree}
\label{table:confusion:decision}
\begin{tabular}{|c|c|c|c|c|c|c|c|c|c|c|c|c|} \hline

&a&b&c&d&e&f&g&h&i&j&k&l\\ \hline

a&0.8 & & & 0.2 & & & & & & & &  \\ \hline 
  
b&  & 0.8 & & & & & & & & 0.2  & &  \\ \hline
  
c&  &  & 1.0 & & & & & & & & &  \\ \hline
  
d&  &  &  & 0.8 & & & & & 0.2 & & &  \\ \hline  
  
e&  &  &  & & 0.8 & & & & & & 0.2 &  \\ \hline  

f&  &  &  & &  & 0.8 & & 0.2 & & &  &  \\ \hline   
  
g&  &  &  & &  &  & 1.0 &  & & &  &  \\ \hline     
  
h&  &  &  & &  &  &  & 1.0 & & &  &  \\ \hline    
  
i&  &  &  & 0.2 &  &  &  &  & 0.8 & &  &  \\ \hline    
  
j&  &  &  &  &  &  &  &  & & 1.0 &  &  \\ \hline

k&  &  &  &  &  &  &  &  & & & 1.0   &  \\ \hline     

l&  &  &  &  &  &  &  &  & & & 0.2  & 0.8 \\ \hline     

\end{tabular}
\end{table}

%%%%%%%%%%%%%%%%%%%%%%%%%%%%%%%%%%%%%%%%%%%%%%%%%%%%%%%%%%%%%%%%%%%%%%%%%%%%%%%%%%%%%%%%
\section{Discussion}
Results of the preliminary studies suggest that gaze-assisted authentication is not vulnerable to shoulder-surfing attacks.
The aspect that there is no way to recognize where on the screen the user is looking, without the assistance of advanced technologies, makes gaze-assisted authentication a strong approach to preventing shoulder-surfing attacks.
During the user studies, no user had an issue with following the moving shapes, and this is due the natural ability of the humans to follow moving objects, and it does not induce any cognitive load.
Most importantly, our study suggests that gaze-assisted authentication is even non-vulnerable to brute force attacks since every individual has unique pattern of eye movements.
This is the reason why attempts to break into the system by generating paths automatically by a computer may have limited success when compared to hacking a PIN based password.

Regarding the system accuracy, from the user studies we found that template matching algorithm has a higher accuracy (95\%) than the decision tree algorithm (90.2\%).
This result is not surprising since the template matching algorithm matches the user's scan-path against all the template paths, and finds the template path that is at the least Euclidean distance.
On the other hand, the decision tree algorithm extracts the unique features of the template paths and their associated classes.
Any candidate path whose features match closely with a given feature set in the model, the class of the matching template is assigned as the class of the candidate path.
Hence, there are chances of errors, resulting in lower accuracy in case of the decision tree algorithm.

Though template matching algorithm is more accurate than the decision tree algorithm, it is relatively slower.
In our studies, we found that the decision tree algorithm is thrice as fast as the template matching algorithm.
Again, this efficiency in speed is expected since the decision tree algorithm checks the features extracted against the decision tree (model) that is built from multiple templates.
In summary, our study suggests that gaze-assisted authentication is a highly secure solution again shoulder-surfing attacks.
For higher accuracy, the template matching algorithm is preferred, but at the cost of higher recognition time.
The decision tree algorithm on the other hand provides speed efficiency, but the accuracy would be lower.

%%%%%%%%%%%%%%%%%%%%%%%%%%%%%%%%%%%%%%%%%%%%%%%%%%%%%%%%%%%%%%%%%%%%%%%%%%%%%%%%%%%%%%%%
\section{Conclusion}
In this work we presented a gaze-assisted authentication system as a potential solution against shoulder-surfing attacks.
Unlike the existing gaze-assisted solutions which are either inaccurate or induce cognitive load on the user, our solution is simple where a user follows moving shapes on the screen to authenticate.
The authentication interface consists of a canvas with twelve shapes on it, and a user chooses three shapes as her password.
Each shape has a pre-defined position on the canvas and traverses along a pre-defined path.
The user's scan-path is matched against the paths traversed by all the shapes, and if all the three scan-paths match then the user is authenticated.
We use both the template matching and decision tree algorithms to match the user's scan-paths.
From our user study involving seven participants, we found that gaze-assisted authentication is a highly secure method to counter shoulder surfing attacks.
Also, using the template matching algorithm is suggested if accuracy is the primary constraint, but the decision tree algorithm is preferred if reducing the time is the main constraint.
%%%%%%%%%%%%%%%%%%%%%%%%%%%%%%%%%%%%%%%%%%%%%%%%%%%%%%%%%%%%%%%%%%%%%%%%%%%%%%%%%%%%%%%%%%%%%%%%%%%
% \balance
\bibliographystyle{abbrv}
\bibliography{paper.bbl}  % sigproc.bib is the name of the Bibliography in this case

\begin{thebibliography}{10}

\bibitem{Bulling:2012:ISG:2207676.2208712}
A.~Bulling, F.~Alt, and A.~Schmidt.
\newblock Increasing the security of gaze-based cued-recall graphical passwords
  using saliency masks.
\newblock In {\em Proceedings of the SIGCHI Conference on Human Factors in
  Computing Systems}, CHI '12, pages 3011--3020, New York, NY, USA, 2012. ACM.

\bibitem{DeLuca:2009:LME:1572532.1572542}
A.~De~Luca, M.~Denzel, and H.~Hussmann.
\newblock Look into my eyes!: Can you guess my password?
\newblock In {\em Proceedings of the 5th Symposium on Usable Privacy and
  Security}, SOUPS '09, pages 7:1--7:12, New York, NY, USA, 2009. ACM.

\bibitem{DeLuca:2007:EEI:1324892.1324932}
A.~De~Luca, R.~Weiss, and H.~Drewes.
\newblock Evaluation of eye-gaze interaction methods for security enhanced
  pin-entry.
\newblock In {\em Proceedings of the 19th Australasian Conference on
  Computer-Human Interaction: Entertaining User Interfaces}, OZCHI '07, pages
  199--202, New York, NY, USA, 2007. ACM.

\bibitem{friedl1997decision}
M.~A. Friedl and C.~E. Brodley.
\newblock Decision tree classification of land cover from remotely sensed data.
\newblock {\em Remote sensing of environment}, 61(3):399--409, 1997.

\bibitem{Kaul:sketch}
P.~Kaul, V.~Rajanna, and T.~Hammond.
\newblock Exploring users' perceived activities in a sketch-based intelligent
  tutoring system through eye movement data.
\newblock In {\em Proceedings of the ACM Symposium on Applied Perception}, SAP
  '16, pages 134--134, New York, NY, USA, 2016. ACM.

\bibitem{Kumar:2007:RSU:1280680.1280683}
M.~Kumar, T.~Garfinkel, D.~Boneh, and T.~Winograd.
\newblock Reducing shoulder-surfing by using gaze-based password entry.
\newblock In {\em Proceedings of the 3rd Symposium on Usable Privacy and
  Security}, SOUPS '07, pages 13--19, New York, NY, USA, 2007. ACM.

\bibitem{Moncur:2007:PAE:1240624.1240758}
W.~Moncur and G.~Lepl\^{a}tre.
\newblock Pictures at the atm: Exploring the usability of multiple graphical
  passwords.
\newblock In {\em Proceedings of the SIGCHI Conference on Human Factors in
  Computing Systems}, CHI '07, pages 887--894, New York, NY, USA, 2007. ACM.

\bibitem{Rajanna:foottyping}
V.~Rajanna.
\newblock Gaze typing through foot-operated wearable device.
\newblock In {\em Proceedings of the 18th International ACM SIGACCESS
  Conference on Computers and Accessibility}, ASSETS '16, pages 345--346, New
  York, NY, USA, 2016. ACM.

\bibitem{Rajanna:gawschi}
V.~Rajanna and T.~Hammond.
\newblock Gawschi: Gaze-augmented, wearable-supplemented computer-human
  interaction.
\newblock In {\em Proceedings of the Ninth Biennial ACM Symposium on Eye
  Tracking Research \& Applications}, ETRA '16, pages 233--236, New York, NY,
  USA, 2016. ACM.

\bibitem{Rajanna:IUI}
V.~D. Rajanna.
\newblock Gaze and foot input: Toward a rich and assistive interaction
  modality.
\newblock In {\em Companion Publication of the 21st International Conference on
  Intelligent User Interfaces}, IUI '16 Companion, pages 126--129, New York,
  NY, USA, 2016. ACM.

\bibitem{Roth:2004:PMR:1030083.1030116}
V.~Roth, K.~Richter, and R.~Freidinger.
\newblock A pin-entry method resilient against shoulder surfing.
\newblock In {\em Proceedings of the 11th ACM Conference on Computer and
  Communications Security}, CCS '04. ACM, 2004.

\bibitem{Wobbrock:2007:GWL:1294211.1294238}
J.~O. Wobbrock, A.~D. Wilson, and Y.~Li.
\newblock Gestures without libraries, toolkits or training: A \$1 recognizer
  for user interface prototypes.
\newblock In {\em Proceedings of the 20th Annual ACM Symposium on User
  Interface Software and Technology}, UIST '07. ACM, 2007.

\end{thebibliography}
\end{document}